\documentclass{jfm}

\usepackage{graphicx,color}
\usepackage{newtxtext}
\usepackage{newtxmath}
\usepackage{natbib}
\usepackage{hyperref}
\hypersetup{
    colorlinks = true,
    urlcolor   = blue,
    citecolor  = black,
}

\newcommand{\GK} [1] {{#1}}
\newcommand{\GKGK} [1] {{#1}}
\newcommand{\GKGKGK} [1] {{#1}}
\newcommand{\SM} [1] {{#1}}
\newcommand{\REV} [1] {{#1}}
\newcommand{\REVGK} [1] {{#1}}
\newcommand{\MU} [1] {{#1}}

\newcommand{\REVR} [1] {{#1}}
\newcommand{\REVREV} [1] {{#1}}

\newcommand{\RomanNumeralCaps}[1]
\linenumbers

\title{\GK{The ultimate state of turbulent permeable-channel flow}}

\author{Shingo Motoki\aff{1}
  \corresp{\email{motoki@me.es.osaka-u.ac.jp}},
  Kentaro Tsugawa\aff{1}, Masaki Shimizu\aff{1}
  \and \GK{Genta Kawahara}\aff{1}}

\affiliation{\aff{1}Graduate School of Engineering Science, Osaka University, 1-3 Machikaneyama, Toyonaka, Osaka 560-8531, Japan}

\begin{document}
\maketitle

\begin{abstract}
\GK{Direct numerical simulations have been performed for heat and momentum transfer in internally heated turbulent shear flow \MU{with} constant bulk mean velocity and temperature, $u_{b}$ and $\theta_{b}$, between parallel, isothermal, no-slip and permeable walls.}
The \GK{wall-normal transpiration} velocity on the walls $y=\pm h$ is assumed to be proportional to the local pressure fluctuations, i.e. $v=\pm \beta p/\rho$ (Jim\'enez {\it et al.}, {\it J. Fluid Mech.}, vol. 442, 2001, pp.89--117).
%Here $\rho$ is \GK{the} mass density of the fluid, and the property of the permeable wall is given by the \GK{dimensionless parameter $\beta u_{b}$}.
The temperature is \GK{supposed to be} a passive scalar, and the Prandtl number is set to unity.
\REVR{\REVGK{Turbulent heat and momentum transfer in permeable-channel flow for $\beta u_{b}=0.5$ has been found to exhibit distinct states depending on the Reynolds number $Re_b=2h u_b/\nu$.
At $Re_{b}\lesssim 10^4$, the classical Blasius law of the friction coefficient and its similarity to the Stanton number, $St\approx c_{f}\sim Re_{b}^{-1/4}$, are observed, whereas at $Re_{b}\gtrsim 10^4$, the so-called ultimate scaling, $St\sim Re_b^0$ and $c_{f}\sim Re_b^0$, is found.}
\REV{
%In the permeable channel for $\beta u_{b}=0.5$ we have found the transition of the scaling of the Stanton number $St$ and the friction coeficient $c_{f}$ from the classical Blasius law $St\approx c_{f}\sim Re_{b}^{-1/4}$ to $St\sim Re_{b}^{0}$ and $c_{f}\sim Re_b^0$ representing the so-called ultimate state.
The ultimate state is attributed to the appearance of large-scale intense spanwise rolls with the length scale of $O(h)$ arising from the Kelvin--Helmholtz type \REVGK{of shear-layer} instability over the permeable walls.
\REVGK{The large-scale rolls can induce large-amplitude velocity fluctuations of $O(u_b)$ as in free shear layers, so that the Taylor dissipation law $\epsilon\sim u_{b}^{3}/h$ (or equivalently $c_{f}\sim Re_b^0$) holds.
In spite of strong turbulence promotion there is no flow separation, and thus large-amplitude temperature fluctuations of $O(\theta_b)$ can also be induced similarly.}
As a consequence, the ultimate heat transfer is achieved, i.e., a wall heat flux scales with $u_{b}\theta_{b}$ (or equivalently $St\sim Re_b^0$) independent of thermal diffusivity, although the heat transfer on the walls is dominated by thermal conduction.}}
\end{abstract}


\section{Introduction}\label{sec:introduction}
\REVGK{One of the major issues in engineering and geophysics is to understand the effects of wall surface properties on heat and momentum transfer in turbulent shear flows.
Turbulent} \GK{flows} over rough walls \GK{have} been extensively investigated \GK{experimentally and numerically}~\citep[\GK{see}][]{Jimenez2004}.
%In turbulent \GK{flows}, 
\REVGK{Surface} roughness \GK{on a wall usually} increases the drag \GK{thereon} in comparison to a smooth wall.
\GK{In the fully rough regime at high Reynolds numbers $Re$, the friction coefficient $c_{f}$ can be independent of $Re$ as seen in the Moody diagram \citep{Moody1944}.}
The scaling \GK{$c_{f}\sim Re^0$} corresponds to \GK{the Taylor dissipation law implying \MU{that the energy dissipation is} independent} of the kinematic viscosity \GK{$\nu$}.
\MU{It is well known that in wall turbulence there exists a similarity between heat and momentum transfer, which can be empirically expressed as a relation between the Stanton number $St$ (i.e.\ a dimensionless wall heat flux) and the friction coefficient $c_f$, viz.\ $St\sim Pr^{-2/3}c_{f}$ \citep{Chilton1934}, where $Pr$ is the Prandtl number.}
\GK{In rough-wall flows, however, $St$ decreases as $Re$ increases even in the fully rough regime where $c_{f}\sim Re^0$} \citep{Dipprey1963,Webb1971}.
\GK{This dissimilarity is a consequence of flow separation from roughness elements.
In the fully rough regime at high $Re$ (for $Pr\sim 1$), the viscous sublayer separates from the roughness elements to yield pressure drag on the rough wall, whereas the thin thermal conduction layer without any vortices is stuck to the rough surface \citep{MacDonald2019a}.

The scaling $St\sim Re^0$ in forced convection means that \MU{the} wall heat flux is independent of the thermal diffusivity $\kappa$.
It relates to the well known ultimate scaling $Nu\sim Pr^{1/2}Ra^{1/2}$ (also implying the $\kappa$-independent wall heat flux) suggested by \cite{Spiegel1963,Kraichnan1962} for turbulent thermal convection at extremely high $Ra$, where $Nu$ is the \SM{Nusselt number}, and $Ra$ is the Rayleigh number.
The ultimate \GK{scaling} has been \MU{intensely disputed} in turbulent Rayleigh--B\'enard convection \citep[\GK{see}][]{Ahlers2009,Chilla2012,Roche2020}.}
In thermal convection, it has been found that \REVGK{wall roughness} %transiently 
yields the scaling $Nu\sim Pr^{1/2}Ra^{1/2}$ in the limited range of $Ra$ where the thermal conduction layer thickness is comparable to the size of \MU{the} roughness elements \citep{Zhu2017,Zhu2019,Macdonald2019b}.
\REVR{It is still an open question whether or not the \REV{ultimate scaling can \MU{actually} be achieved \REVGK{at high $Re$ or $Ra$}
by introducing \MU{a specifically engineered surface}} \GK{in forced or thermal convection}.}
%by introducing sophisticated roughness geometries.

Recently, \cite{Kawano2021} have found that the ultimate heat transfer $Nu\sim Pr^{1/2}Ra^{1/2}$ can be achieved in turbulent thermal convection between permeable walls.
In \GK{their study}, the \GK{wall-normal} transpiration velocity on the wall is assumed to be proportional to the local pressure fluctuations.
This \GK{permeable} boundary condition was originally introduced by \cite{Jimenez2001} \GK{to mimic} a Darcy-type porous wall with a constant-pressure plenum chamber \REVGK{underneath}.
\GK{They} have investigated turbulent momentum transfer in \GK{permeable-channel} flow, and found that the \REVR{wall-transpiration} leads to large-scale spanwise rolls over the permeable wall, significantly \GK{enhancing} momentum transfer.
\GK{By linear stability analyses,
\cite{Jimenez2001} have clarified that the formation of the large-scale spanwise rolls originates from the Kelvin--Helmholtz type \REVGK{of} shear-layer instability over the \SM{permeable} wall.}
Such large-scale \GK{turbulence} structures have been observed \GK{numerically and experimentally} in shear flows over porous media \citep[see e.g.][]{Suga2018,Nishiyama2020}.

In \GK{the present} study, we investigate the scaling properties \GK{of heat and momentum transfer} in turbulent \MU{channel flow with permeable walls} and report that the \REVR{wall-transpiration} \GK{can \SM{bring about} the ultimate state represented by the \MU{viscosity-independent dissipation} $c_{f}\sim Re^0$ as well as the \MU{diffusivity-independent heat flux} $St\sim Re^0$}.
%, and then discuss the turbulent structures and statistics quite distinct from that observed in turbulent channel flows between impermeable walls and rough walls.
%
\vspace*{-2mm}
\section{Governing equations and numerical \GK{simulations}}\label{sec:formulation}
Let us consider \GK{turbulent heat and \SM{momentum} transfer in internally heated shear flow between parallel, isothermal, no-slip and permeable walls.}
The coordinates, $x$, $y$ and $z$ (or $x_{1}$, $x_{2}$ and $x_{3}$) are used for the representation of the streamwise, the wall-normal and the spanwise directions, respectively.
The origin \REVGK{of the coordinate system} is on the midplane \GK{between} the two walls positioned at $y=\pm h$.
The corresponding components of the velocity $\mbox{\boldmath$u$}(\mbox{\boldmath$x$},t)$ are given by $u,v$ and $w$ (or $u_{1},u_{2}$ and $u_{3}$), respectively.
The temperature $\theta(\mbox{\boldmath$x$},t)$ is \GK{supposed to be} a passive scalar.
The governing equations are the Navier--Stokes equations for the divergence-free velocity and the \GK{energy} equation for the temperature,
\begin{eqnarray}
\label{eq:co}
\displaystyle
\nabla\cdot\textit{\textbf{u}}&=&0,\\
\label{eq:ns}
\displaystyle
\frac{\partial \textit{\textbf{u}}}{\partial t}+(\textit{\textbf{u}}\cdot\nabla)\textit{\textbf{u}}&=&\GK{\nu\nabla^{2}\textit{\textbf{u}}+f\mbox{\boldmath$e$}_{x}-\frac{1}{\rho}\nabla p},\\
\label{eq:ad}
\displaystyle
\frac{\partial \theta}{\partial t}+(\textit{\textbf{u}}\cdot\nabla)\theta&=&\kappa\nabla^{2}\theta+\frac{q}{\rho c_{p}},
\end{eqnarray}
where $p(\mbox{\boldmath$x$},t)$ is the fluctuating pressure with respect to the driving pressure $P(x,t)$, 
\REVGK{and}
$\rho,\nu,\kappa$ and $c_{p}$ are the mass density, the kinematic viscosity, the thermal diffusivity and the \GK{specific heat at constant pressure} of the fluid, respectively.
\GKGK{Here,}
vector $\mbox{\boldmath$e$}_{x}$ is a unit vector in the streamwise direction, \GKGK{and}
\REVGK{$f(t)$ ($=-\rho^{-1}\partial P/\partial x>0$)} and \GK{$q(t)$ ($>0$)} are the spatially uniform driving force and internal heat source to maintain constant bulk mean velocity and temperature, $u_{b}$ and $\theta_{b}$, respectively.
\REVR{\REVGK{The momentum equation (\ref{eq:ns}) and the energy equation (\ref{eq:ad}) are similar in the sense that they have the corresponding terms except for the (rightmost) pressure fluctuation term in (\ref{eq:ns}).
As a consequence, we can observe similarity between momentum and heat transfer in turbulent shear flows, although strong local pressure fluctuations occasionally bring about significant dissimilarity.}}
The velocity and temperature fields are supposed to be periodic in the $x$- and $z$-directions with the periods, $L_{x}$ and $L_{z}$.
\GK{On the permeable wall the wall-normal velocity $v$ is assumed to be proportional to the local pressure fluctuation $p$ \citep{Jimenez2001,Kawano2021}.
	We impose the no-slip, permeable and isothermal conditions,}
\begin{eqnarray}
\label{eq:bc}
u(y=\pm h)=w(y=\pm h)=0,\hspace{1em}v(y=\pm h)=\pm\beta\frac{p\GKGKGK{(y=\pm h)}}{\rho};\hspace{1em}\theta(y=\pm h)=0,
\end{eqnarray}
\GK{on the walls}, where \GK{$\beta$ ($\ge 0$)} represents the \REVR{\MU{`permeability' parameter}}, and the impermeable conditions $v(y=\pm h)=0$ are recovered for $\beta=0$, while $\beta\rightarrow\infty$ implies zero pressure fluctuations and an unconstrained wall-normal velocity.
Note that the pressure fluctuation with zero mean instantaneously ensures a zero net mass flux through the permeable wall.
\REVREV{We anticipate the no-slip and permeable conditions (\ref{eq:bc}) on a wall perforated with many fine holes connected to an adjacent constant-pressure plenum chamber (see the last paragraph in \S~\ref{sec:results} for the realistic configuration).
Actually, we have confirmed that the mean and fluctuation velocities over the no-slip permeable wall are in good agreement with those observed experimentally \citep{Suga2010} and numerically \citep{Breugem2006} over a porous wall.}

The flow is characterised by the bulk Reynolds number $Re_{b}=2hu_{b}/\nu$, the Prandtl number $Pr=\nu/\kappa$ and the dimensionless \REVR{\MU{permeability parameter}} $\beta u_{b}$.
The wall heat flux $q_{w}$ and the wall \GKGK{shear stress} $\tau_{w}$ are \REVGK{respectively} quantified by the Stanton number $St$ and the friction coefficient \GK{$c_{f}$} defined as
\begin{eqnarray}
\label{eq:stcf}
\displaystyle
\GK{
	St\equiv\frac{2q_{w}}{\rho c_{p}u_{b}\theta_{b}}=2\frac{u_{\tau}\theta_{\tau}}{u_{b}\theta_{b}},\hspace{1em}
	c_{f}\equiv\frac{2\tau_{w}}{\rho u_{b}^{2}}=2\frac{u_{\tau}^{2}}{u_{b}^{2}},
}
\end{eqnarray}
where \GK{$u_{\tau}={(\mp\nu{{\rm d}{\left< u \right>}_{xzt}/{\rm d}y}|_{y=\pm h})}^{1/2}$} and \GK{$\theta_{\tau}=\mp(\kappa/u_{\tau}){\rm d}{\left< \theta \right>}_{xzt}/{\rm d}y|_{y=\pm h}$} are the friction velocity and the friction temperature, respectively.
\GK{Hereafter}, ${\left< \cdot \right>}_{xzt}$ and ${\left< \cdot \right>}_{xyzt}$ represent \REVGK{a} \GK{plane-time} ($xzt$-) average and \REVGK{a} \GK{volume-time} ($xyzt$-) average, respectively.

We conduct \GK{direct numerical simulations (DNS)} for turbulent heat and \GK{momentum} transfer in \GK{shear} flow between permeable walls.
The present DNS code is based on the \GK{one} developed for turbulent thermal convection between permeable walls \citep{Kawano2021}.
The governing equations (\ref{eq:co})--(\ref{eq:ad}) are discretised employing the spectral Galerkin method based on the Fourier--Chebyshev expansions.
Time advancement is performed with \GK{the aid of} \MU{the implicit Euler scheme for the diffusion terms and a third-order Runge--Kutta scheme otherwise.}
In this paper, we present results obtained \MU{for $\beta u_{b}=0$ (referred to as the impermeable case), $\beta u_{b}=0.3$ (referred to as the less-permeable case) and $\beta u_{b}=0.5$ (referred to as the permeable case), in all cases for $Pr=1$}.
The simulations are carried out \GK{at $Re_b=4\times 10^3$--$4\times 10^4$} \MU{in periodic computational boxes of size $(L_{x},L_{z})=(2\pi h,\pi h)$}.
The spatial grid spacings are less than $10$ wall units in all the \REVGK{three} directions, and the \GK{data are accumulated for the duration of more than $30$ wall units at $\beta u_b=0.5$.}

\section{Heat flux, shear stress and energy budget}\label{sec:budget}
In this section, we \GKGK{show} the total heat flux, the total shear stress and the total energy budget in \GK{internally heated shear} flow between permeable walls.
\GK{We} decompose the velocity and temperature into \GK{an} $xzt$-average and a fluctuation about it as $\mbox{\boldmath$u$}={\left< \mbox{\boldmath$u$} \right>}_{xzt}+\mbox{\boldmath$u$}'$ and $\theta={\left< \theta \right>}_{xzt}+\theta'$.
Substituting \REVGK{the decompositions into} (\ref{eq:ns}) and (\ref{eq:ad}), integrating their $xzt$-averages with respect to $y$, and \GK{supposing} that the flow is statistically \GK{stationary}, we obtain the total heat flux and the total shear stress, \GK{respectively,}
\begin{eqnarray}
\label{eq:flux}
\displaystyle
\kappa\frac{{\rm d}{\left< \theta \right>}_{xzt}}{{\rm d}y}-{\left< \theta'v'  \right>}_{xzt}&=&-\frac{{\left< q \right>}_{t}}{\rho \GK{c_{p}}}(y+h)+\GK{u_{\tau}\theta_{\tau}},\\
\label{eq:shear}
\displaystyle
\nu\frac{{\rm d}{\left< u \right>}_{xzt}}{{\rm d}y}-{\left< u'v' \right>}_{xzt}&=&-{\left< f \right>}_{t}(y+h)+\GK{u_{\tau}^{2}},
\end{eqnarray}
\GK{where ${\left< \cdot \right>}_{t}$ stands for \REVGK{a} time average.}
Note that the turbulent heat flux ${\left< \theta'v' \right>}_{xzt}$ and the Reynolds shear stress \GK{$-{\left< u'v' \right>}_{xzt}$} \GKGK{have vanished} on the walls ($y=\pm h$) even in \REVGK{a} permeable case due to the \GKGK{isothermal and no-slip} conditions.
\REVR{\REVGK{Recalling $c_f$ in (\ref{eq:stcf}) and using (\ref{eq:shear}) for $y=h$, we have the balance between the friction drag and the driving force,}}
\begin{eqnarray}
\label{eq:cf-fu}
\REVR{\REVGK{c_f u_b^2=2{\left< f \right>}_{t}h.}}
\end{eqnarray}

\REVR{By taking the $xyzt$-average of \MU{the} inner product of the Navier--Stokes equation (\ref{eq:ns}) with the velocity $\mbox{\boldmath$u$}$ and taking account of the boundary conditions (\ref{eq:bc}), we obtain the total energy budget equation}
\begin{eqnarray}
\label{eq:budget}
%\displaystyle
\REVR{\REVGK{
\epsilon+\frac{1}{2h\beta}\left(\left. {\left< v^{2} \right>}_{xzt}\right|_{y=h}+\left.{\left< v^{2} \right>}_{xzt}\right|_{y=-h} \right)+\frac{1}{4h}{\left[ {\left< v^{3} \right>}_{xzt} \right]}_{y=-h}^{y=h}
={\left< f \right>}_{t}u_{b}
=\frac{c_f}{2}\frac{u_b^3}{h}},}
\end{eqnarray}
\REVR{where $\epsilon=(\nu/2){\langle {\left( \partial u_{i}/\partial x_{j}+\partial u_{j}/\partial x_{i} \right)}^{2} \rangle}_{xyzt}$ is \MU{the} total energy dissipation rate per unit mass,
\REVGK{and \MU{where} we have used ${\langle fu \rangle}_{xyzt}={\langle f{\langle u \rangle_{xyz}}\rangle}_{t}={\langle f \rangle}_{t}u_{b}$}.
\REVGK{The rightmost equality is given by (\ref{eq:cf-fu}).}
\MU{From (\ref{eq:budget}) it can be seen that the effect of the wall-transpiration appears in two terms of the total energy budget.
The second term on the left-hand side denotes the work done by pressure at the permeable walls.}}
This term is strictly non-negative, being an energy sink.
The third term represents outflow kinetic energy across the permeable walls.
In the present DNS we have confirmed that the second term is at most $1\%$ of $\epsilon$ whereas the third term is less than $0.01\%$ of $\epsilon$.
\GK{\MU{Hence, it} turns out that the introduction of the \REVR{wall-transpiration} does not bring about any extra energy inputs.}

\begin{figure}
	\centering
	\begin{minipage}{.9\linewidth}
		\includegraphics[clip,width=\linewidth]{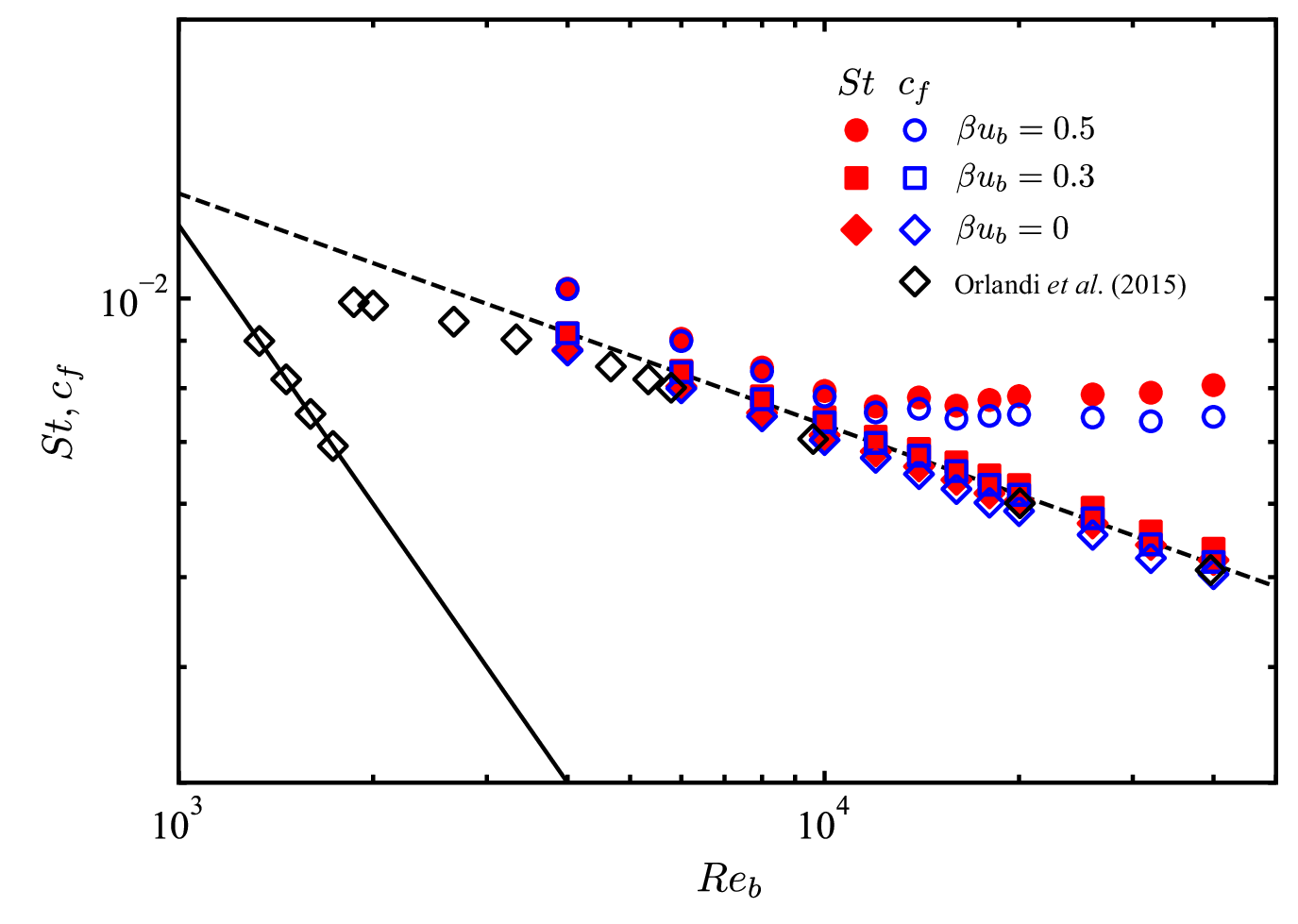}
	\end{minipage}
	\caption{\REVR{\REV{Stanton number $St$ and friction coefficient $c_{f}$ as a function of bulk Reynolds number $Re_{b}$ in permeable- and impermeable-channel flows for Prandtl number $Pr=1$.
    \REVGK{The filled red and open blue symbols represent $St$ and $c_{f}$, respectively.
    The permeable case \MU{($\beta u_{b}=0$)}, the less-permeable case \MU{($\beta u_{b}=0.3$)}, and the permeable case \MU{($\beta u_{b}=0.5$)} are indicated by diamonds, squares, and circles, respectively.}
    The open black diamonds denote $c_{f}$ in the DNS taken from \cite{Orlandi2015}.
    The solid and dashed lines indicate $c_{f}=12 Re_{b}^{-1}$ for laminar flow and the empirical formula $c_{f}=0.073Re_{b}^{-1/4}$ \citep{Dean1978} for turbulent flow, respectively.}}
	\label{fig:stcf-reb}}
\end{figure}

\begin{figure}
	\centering
		\includegraphics[clip,width=.97\linewidth]{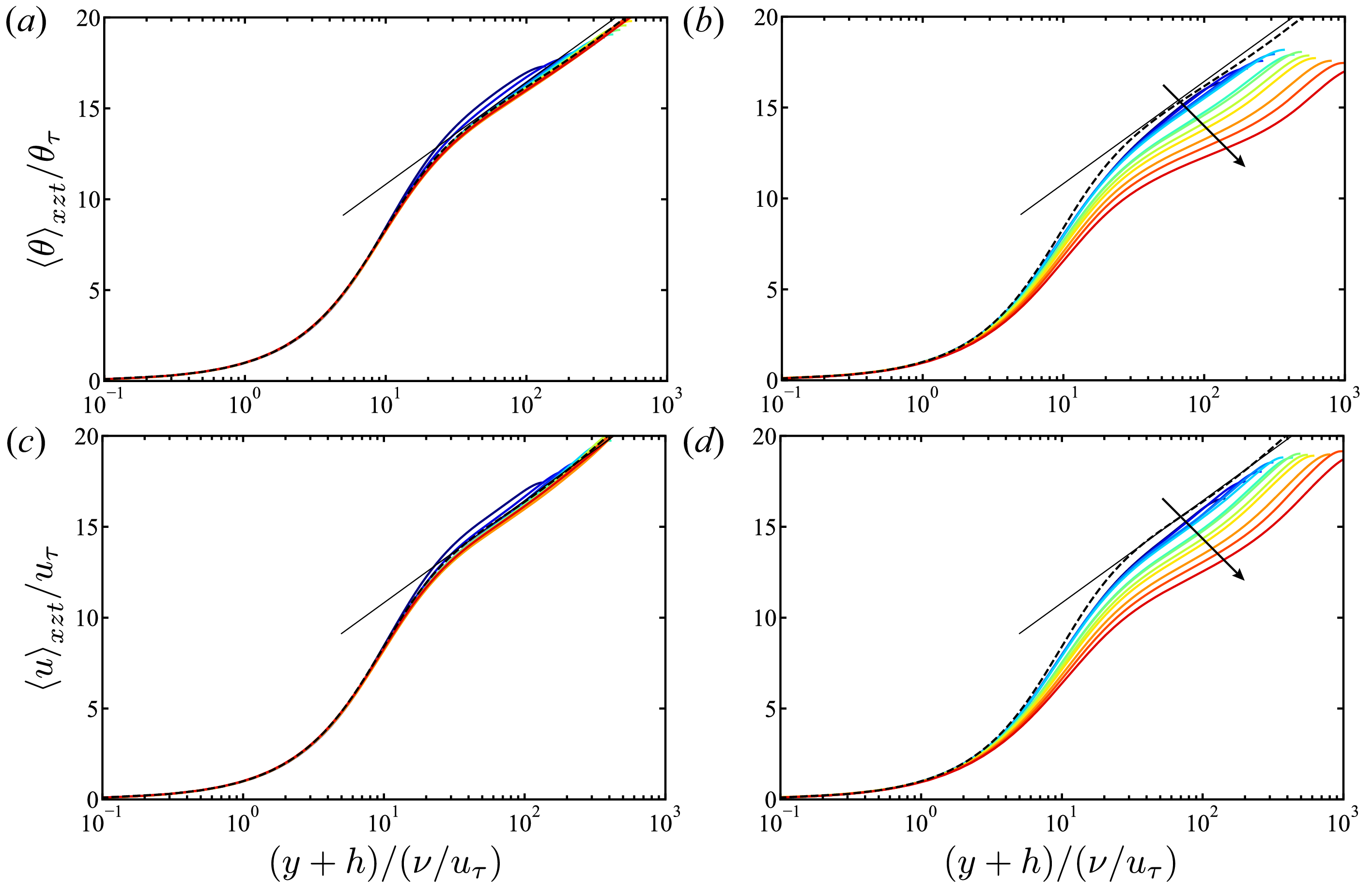}
	\caption{Mean temperature and velocity \REVR{\REVGK{respectively normalised by $\theta_{\tau}$ and $u_{\tau}$}} as a function of the distance to the lower wall $(y+h)/(\nu/u_{\tau})$ in (\textit{a,c}) the \GK{less}-permeable case \MU{($\beta u_{b}=0.3$)} and (\textit{b,d}) the \GK{permeable} case \MU{($\beta u_{b}=0.5$)} at $4\times10^{3}\le Re_{b}\le4\times10^{4}$
		\GK{for $Pr=1$}.
		\GK{The Reynolds number $Re_b$ increases in the direction of the arrows.}
		The dashed lines denote the DNS data \citep{Pirozzoli2016} in \GK{impermeable}-channel flow at $Re_{b}=3.96\times10^{4}$.
		The solid lines \GK{represent the logarithmic law} ${\left< \theta \right>}_{xzt}/\theta_{\tau}={\left< u \right>}_{xzt}/u_{\tau}=(1/0.41)\ln{[(y+h)/(\nu/u_{\tau})]}+5.2$.
		\label{fig:mean_ut_vel_te}}
\end{figure}

\begin{figure}
	\centering
		\includegraphics[clip,width=.97\linewidth]{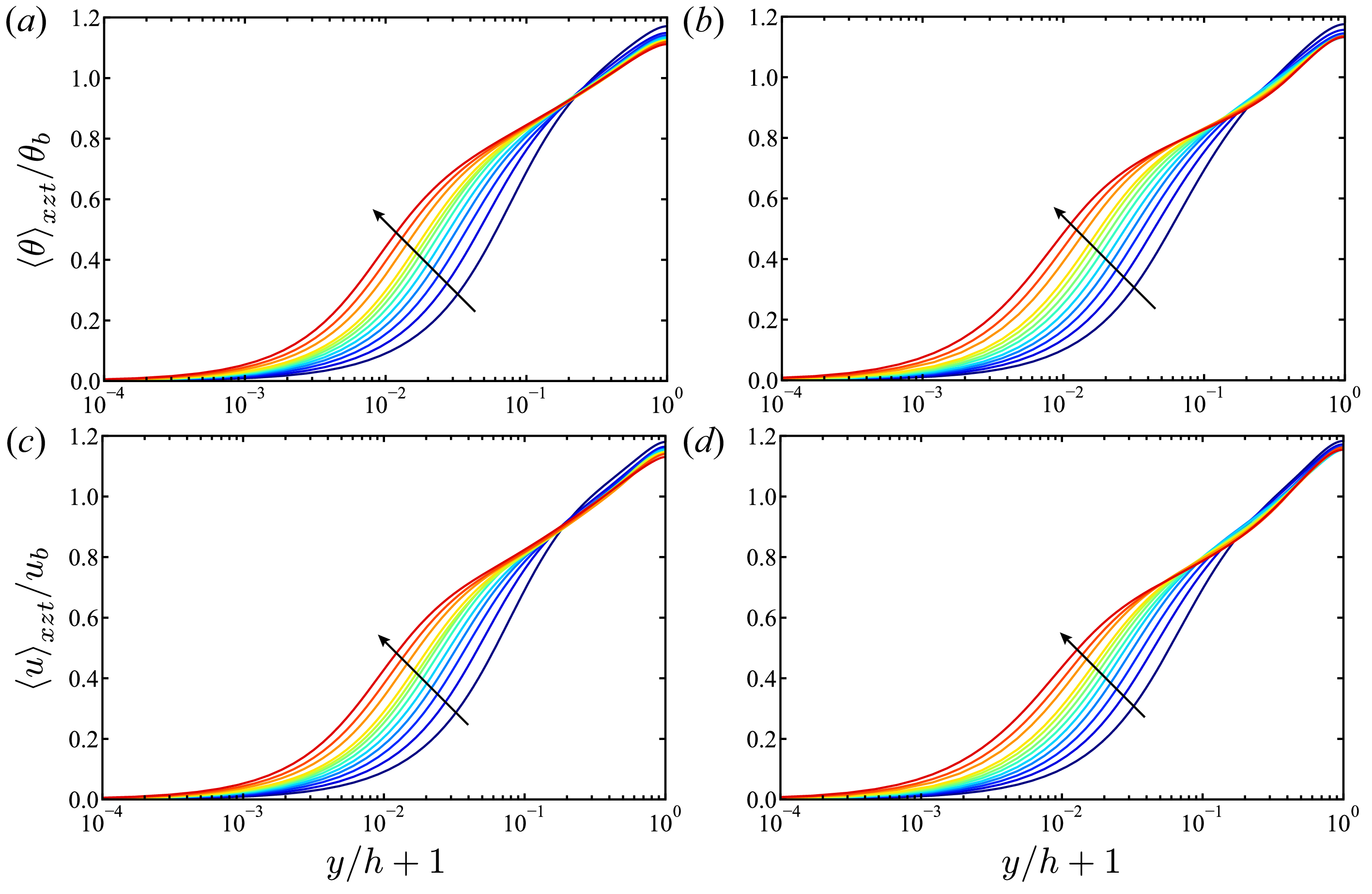}
	    \caption{\REVR{\REVGK{The same as figure~\ref{fig:mean_ut_vel_te} but for mean temperature and velocity respectively normalised by $\theta_b$ and $u_b$ as a function of $y/h+1$.}}
		\label{fig:mean_ub_vel_te}}
\end{figure}

\section{Results \GK{and discussion}}\label{sec:results}

%subsection{$St,C_{f}$-$Re_{b}$ scaling}\label{sec:stcf}
Let us first \REVGK{examine} the \GK{effects} of the \REVR{wall-transpiration} on the Stanton number $St$ and the friction coefficient \GK{$c_{f}$}.
Figure \ref{fig:stcf-reb} shows $St$ and \GK{$c_{f}$} as a function of $Re_{b}$.
In the impermeable case \MU{($\beta u_{b}=0$)} the present DNS data are in good agreement with \GKGK{the numerical result \REVGK{obtained} by} \cite{Orlandi2015} for \REVGK{impermeable-channel} flow in larger \GK{periodic} domains $(L_{x},L_{z})=(12\pi h,4\pi h)$ at $Re_{b}<10^{4}$ and $(L_{x},L_{z})=(6\pi h,2\pi h)$ at $Re_{b}>10^{4}$.
As the \REVR{wall-transpiration} increases \GK{from $\beta u_{b}=0$ to $\beta u_{b}=0.5$}, not only the momentum transfer but the heat transfer are enhanced over the entire \GK{range of} $Re_{b}$.
In the \GK{less}-permeable case \MU{($\beta u_{b}=0.3$)}, $St$ and \GK{$c_{f}$} can be seen to scale with $Re_{b}^{-1/4}$ \REV{at $Re_{b}=4\times10^{3}$--$4\times10^{4}$} \REVGK{as in the impermeable case}, and they exhibit \GK{close} similarity between heat and momentum transfer, i.e. $St\approx \GK{c_{f}}$.
In the \GK{permeable} case \MU{($\beta u_{b}=0.5$)}, on the other hand, the ultimate \GK{state, $St\sim Re_b^0$} and \GK{$c_{f}\sim Re_b^0$}, can be observed \REV{at $Re_{b}\gtrsim 10^{4}$, whereas the classical similar scaling $St\approx c_{f}\sim Re_{b}^{-1/4}$ appear at lower $Re_{b}$ as in the impermeable and less-permeable cases}.

%\subsection{Mean temperature and velocity}\label{sec:mean}
\REVR{Next, we present \REVGK{remarkable differences} in the mean temperature and velocity profiles \REVGK{between} the \GK{less-permeable case \MU{($\beta u_{b}=0.3$)} and the permeable case \MU{($\beta u_{b}=0.5$})}.
The mean temperature and velocity profiles \REVGK{respectively} normalised by the friction temperature $\theta_{\tau}$ and the friction velocity $u_{\tau}$ are shown \REVGK{as a function of the distance to the lower wall, $(y+h)/(\nu/u_{\tau})$, at $Re_b=4\times 10^3$--$4\times 10^4$} in figure~\ref{fig:mean_ut_vel_te}.
\REVGK{In the less-permeable case (figure \ref{fig:mean_ut_vel_te}\textit{a,c}), the normalised mean temperature and velocity, $\left<\theta\right>_{xzt}/\theta_\tau$ and $\left<u\right>_{xzt}/u_\tau$, as a function of $(y+h)/(\nu/u_{\tau})$ do not depend on the Reynolds number, exhibiting the Prandtl wall law (including the logarithmic layer with the prefactor $1/0.41$ and intercept $5.2$ at $(y+h)/(\nu/u_{\tau})\gtrsim 30$) commonly observed in wall turbulence.}
In the \GK{permeable} case (figure \ref{fig:mean_ut_vel_te}\textit{b,d}) \REVGK{at higher Reynolds numbers $Re_{b}\gtrsim 10^{4}$}, on the other hand, the normalised mean temperature and velocity \REVGK{profiles represent significant $Re_b$-dependence at $(y+h)/(\nu/u_{\tau})\gtrsim 10^0$.}
\REVGK{As shown in figure~\ref{fig:mean_ub_vel_te}(\textit{b,d}), the normalised mean temperature and velocity, $\left<\theta\right>_{xzt}/\theta_b$ and $\left<u\right>_{xzt}/u_b$, as a function of $y/h+1$ are nearly independent of $Re_b$ in the bulk region $y/h+1\sim 10^0$ in the permeable case at $Re_{b}\gtrsim 10^{4}$, differing from the known scaling property in wall turbulence (cf. figure~\ref{fig:mean_ub_vel_te}\textit{a,c} in the less-permeable case).
However, since heat and momentum transfer on a permeable wall is dominated by thermal conduction and viscous diffusion due to the isothermal and no-slip boundary conditions as on an impermeable wall, all the profiles in the less-permeable and permeable cases in figure~\ref{fig:mean_ut_vel_te} collapse onto a single line in the linear sublayer $(y+h)/(\nu/u_{\tau})\lesssim 10^0$.}}

\begin{figure}
	\centering
		\includegraphics[clip,width=.97\linewidth]{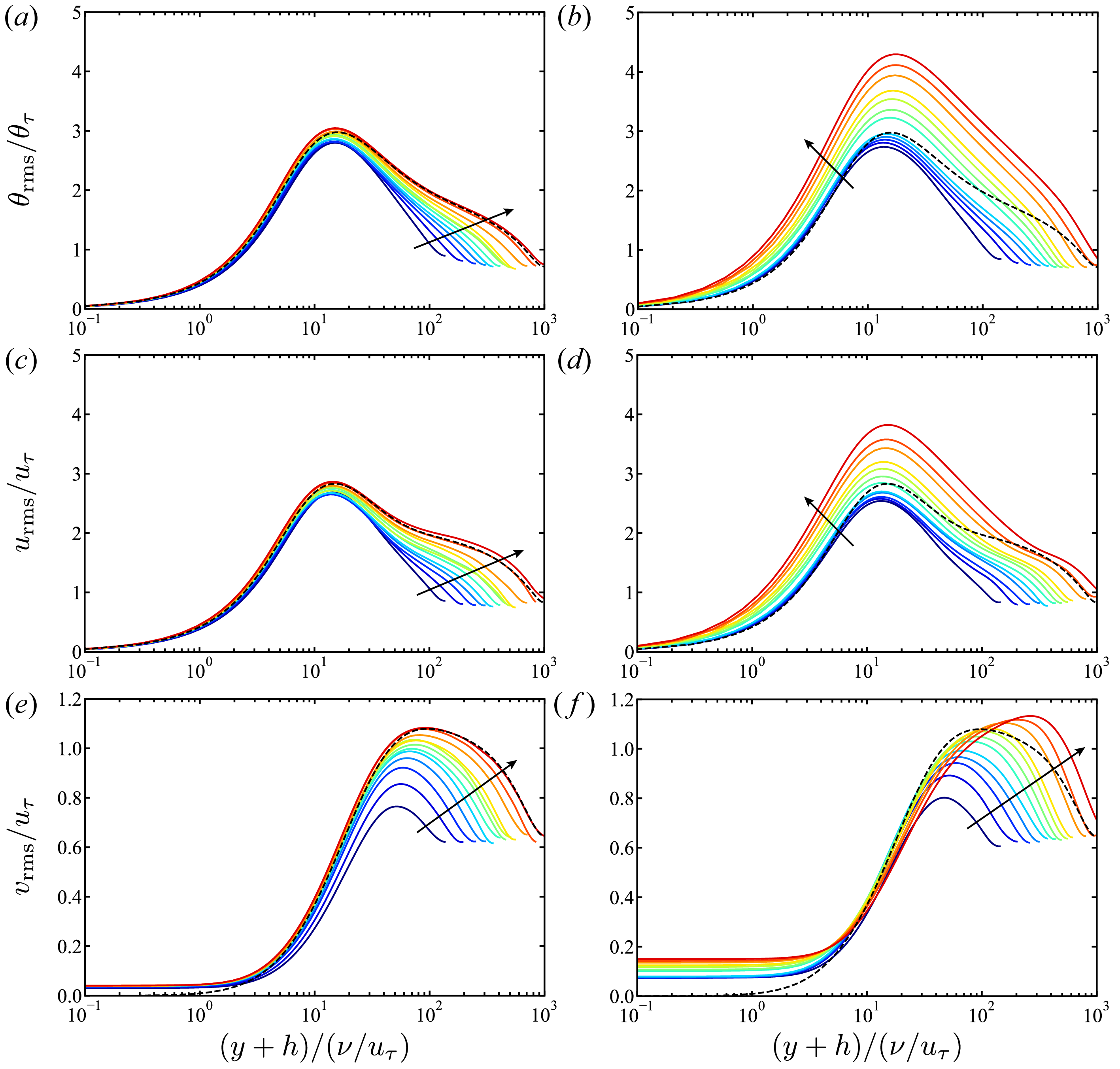}
	\caption{\REVR{\REV{RMS temperature and RMS streamwise and wall-normal \REVGK{velocities respectively} normalised by $\theta_{\tau}$ and $u_{\tau}$ as a function of the distance to the lower wall $(y+h)/(\nu/u_{\tau})$ in (\textit{a,c,e}) the less-permeable case \MU{($\beta u_{b}=0.3$)} and (\textit{b,d,f}) the permeable case \MU{($\beta u_{b}=0.5$)} at $4\times10^{3}\le Re_{b}\le4\times10^{4}$ for $Pr=1$.
	The Reynolds number $Re_b$ increases in the direction of the arrows.
	The dashed lines denote the DNS data \citep{Pirozzoli2016} in impermeable-channel flow at $Re_{b}=3.96\times10^{4}$.}}
		\label{fig:rms_ut_vel_te}}
\end{figure}

\begin{figure}
	\centering
		\includegraphics[clip,width=.97\linewidth]{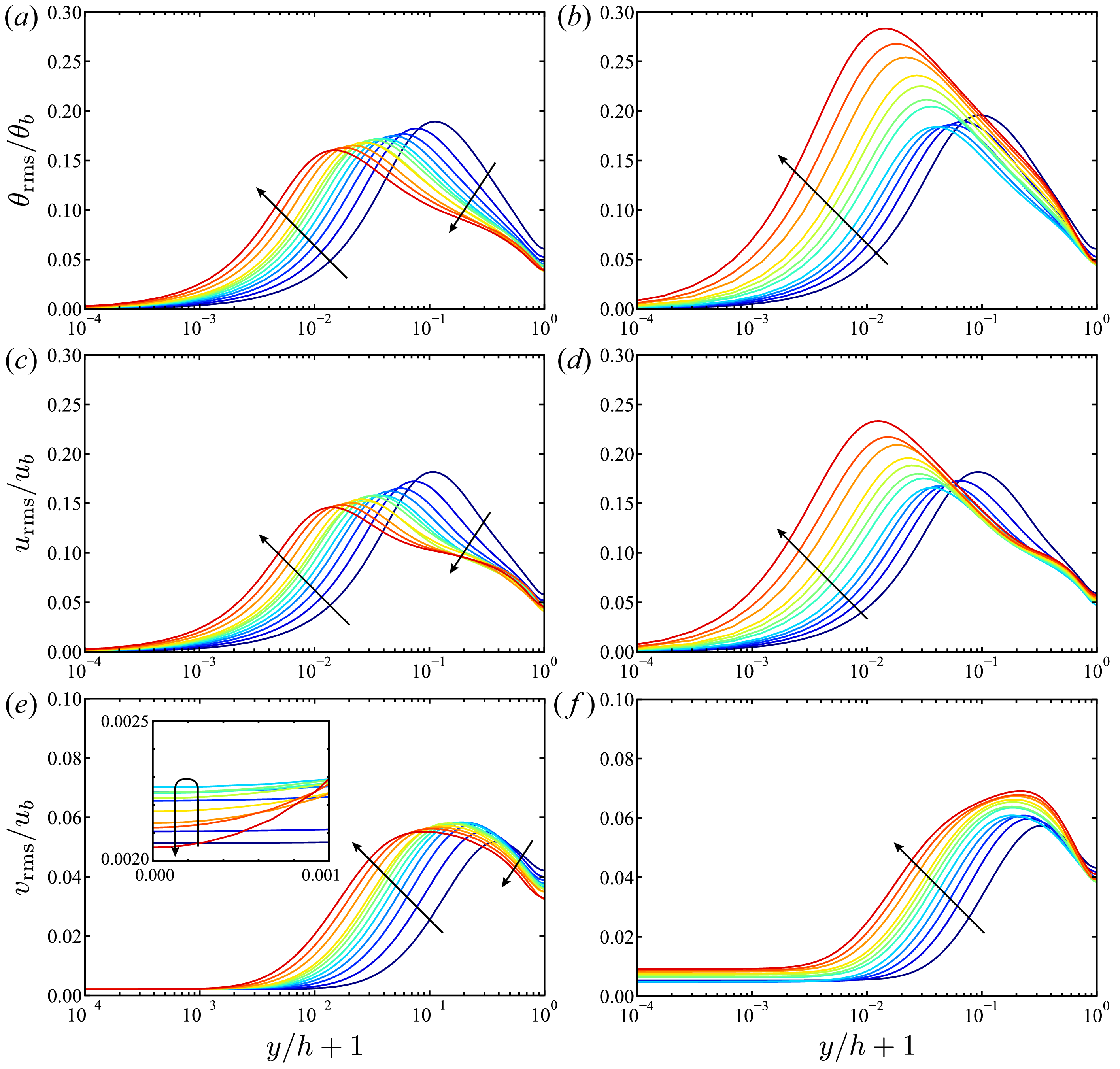}
	\caption{\REVR{\REVGK{The same as figure~\ref{fig:rms_ut_vel_te} but for RMS temperature and velocities respectively normalised by $\theta_b$ and $u_b$ as a function of $y/h+1$. The inset in (\textit{e}) shows $v_{\rm rms}/u_b$ at $0\le y/h+1\le 0.001$.}}
		\label{fig:rms_ub_vel_te}}
\end{figure}

%\subsection{RMS temperature and velocity}\label{sec:rms}
\REVR{\REVGK{Figure~\ref{fig:rms_ut_vel_te} shows the root-mean-square (RMS) temperature $\theta_{\rm rms}={\langle {\theta'}^{2} \rangle}_{xzt}^{1/2}$ and the RMS streamwise and wall-normal velocities, $u_{\rm rms}={\langle {u'}^{2} \rangle}_{xzt}^{1/2}$ and $v_{\rm rms}={\langle v'^{2} \rangle}_{xzt}^{1/2}$, respectively normalised by the friction temperature $\theta_\tau$ and the friction velocity $u_\tau$ as a function of $(y+h)/(\nu/u_\tau)$.}
\REVGK{In the less-permeable case (figure~\ref{fig:rms_ut_vel_te}\textit{a,c,e}), the normalised RMS temperature and velocities, $\theta_{\rm rms}/\theta_\tau$, $u_{\rm rms}/u_\tau$ and $v_{\rm rms}/u_\tau$, as a function of $(y+h)/(\nu/u_\tau)$ are nearly independent of $Re_b$ in the near-wall region, although their $Re_b$-dependence appears remarkably in the bulk region}.
The RMS temperature and \REVGK{streamwise} velocity exhibit almost the same behaviour, \REVGK{suggesting similarity between heat and (streamwise) momentum transfer.
\MU{These properties are consistent with those commonly observed in wall turbulence.}
In the permeable case (figure~\ref{fig:rms_ut_vel_te}\textit{b,d,f}), the similarity between $\theta_{\rm rms}$ and $u_{\rm rms}$ can also be confirmed (see \textit{b,d}) as in the less-permeable case; however, $\theta_{\rm rms}/\theta_\tau$, $u_{\rm rms}/u_\tau$ and $v_{\rm rms}/u_\tau$ as a function of $(y+h)/(\nu/u_{\tau})$ exhibit marked $Re_b$-dependence even in the vicinity of the wall, being distinct from the scaling property observed in wall turbulence.
As shown in figure~\ref{fig:rms_ub_vel_te}(\textit{b,d,f}), the normalised RMS temperature and velocities, $\theta_{\rm rms}/\theta_b$ and $u_{\rm rms}/u_b$ and $v_{\rm rms}/u_b$, as a function of $y/h+1$ are almost independent of $Re_b$ in the bulk region $y/h+1\sim 10^0$ in the permeable case at $Re_{b}\gtrsim 10^{4}$, implying the significant promotion of turbulence of comparable orders with $\theta_b$ and $u_b$ (cf.  figure~\ref{fig:rms_ub_vel_te}\textit{a,c,e} in the less-permeable case).}
%\REV{\REVGK{In the less-permeable case, the near-wall} RMS wall-normal velocity is weak, and \REVGK{$v_{\rm rms}/u_b$} decays with increasing $Re_{b}$ at $Re_{b}\gtrsim10^{4}$ (see \REVGK{the} inset of figure \ref{fig:rms_ub_vel_te}\textit{e}).}
%\MU{Wall-normal transpiration is induced in the permeable case; \GK{however}, the near-wall velocity fluctuation is relatively small in comparison to $u_b$, i.e., $v_{\rm rms}$ near the wall is approximately $1\%$ of $u_{b}$ even at $Re_{b}\gtrsim 10^4$ (figure \ref{fig:rms_ub_vel_te}\textit{f})}.
\REVGK{In short, although conduction- and viscosity-dominated quiescence exists on the wall in the permeable case, intense turbulence is enhanced even in the close vicinity of the permeable wall at higher Reynolds numbers.}}

\begin{figure}
\centering
\includegraphics[clip,width=.76\linewidth]{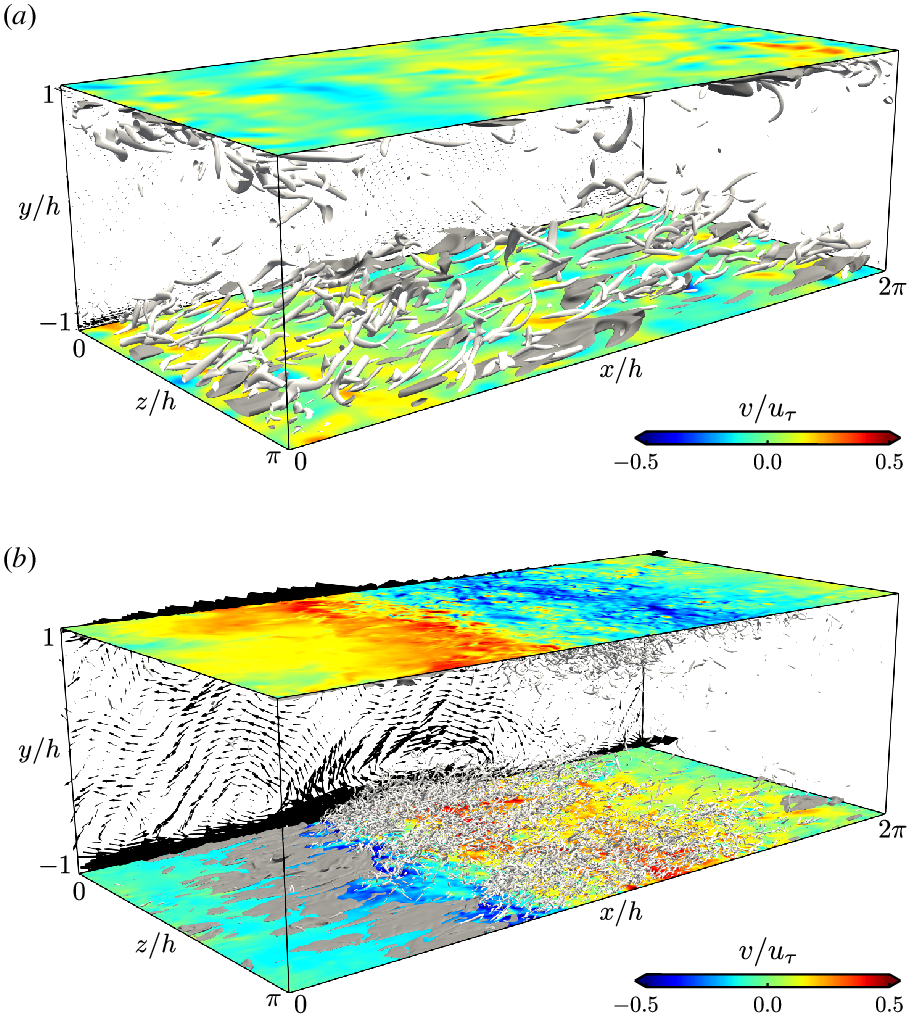}
\caption{Instantaneous flow and thermal structures \GKGK{in} \GK{the permeable \REV{case} ($\beta u_{b}=0.5$) \GKGK{for $Pr=1$} at \REV{(\textit{a}) $Re_{b}=8\times10^{3}$ and (\textit{b}) $Re_{b}=4\times10^{4}$.}}
The \GK{grey} and \GK{dark} grey objects represent the isosurfaces of the positive second invariant of the velocity gradient tensor, (\textit{a}) \GK{$Q/(u_{\tau}/h)^2=2\times10^{3}$} and (\textit{b}) \GK{$Q/(u_{\tau}/h)^2=3\times10^{4}$}\GKGKGK{,} and of the temperature fluctuation (\textit{a}) \GKGK{$\theta'/\theta_{\tau}=4$} and (\textit{b}) \GKGK{$\theta'/\theta_{\tau}=6$}, respectively.
\REVR{\REV{The colour in the top and bottom planes \REVGK{represents the level of} the wall-normal velocity on the walls $y/h=\pm1$.
The vectors in the side plane at $z/h=0$ indicate the spanwise-averaged velocity fluctuations $({\left< u \right>}_{z}-{\left< u \right>}_{xzt},{\left< v \right>}_{z})$.}}
\label{fig:structures}}
\end{figure}

\begin{figure}
	\centering
	\begin{minipage}{\linewidth}
		\includegraphics[clip,width=\linewidth]{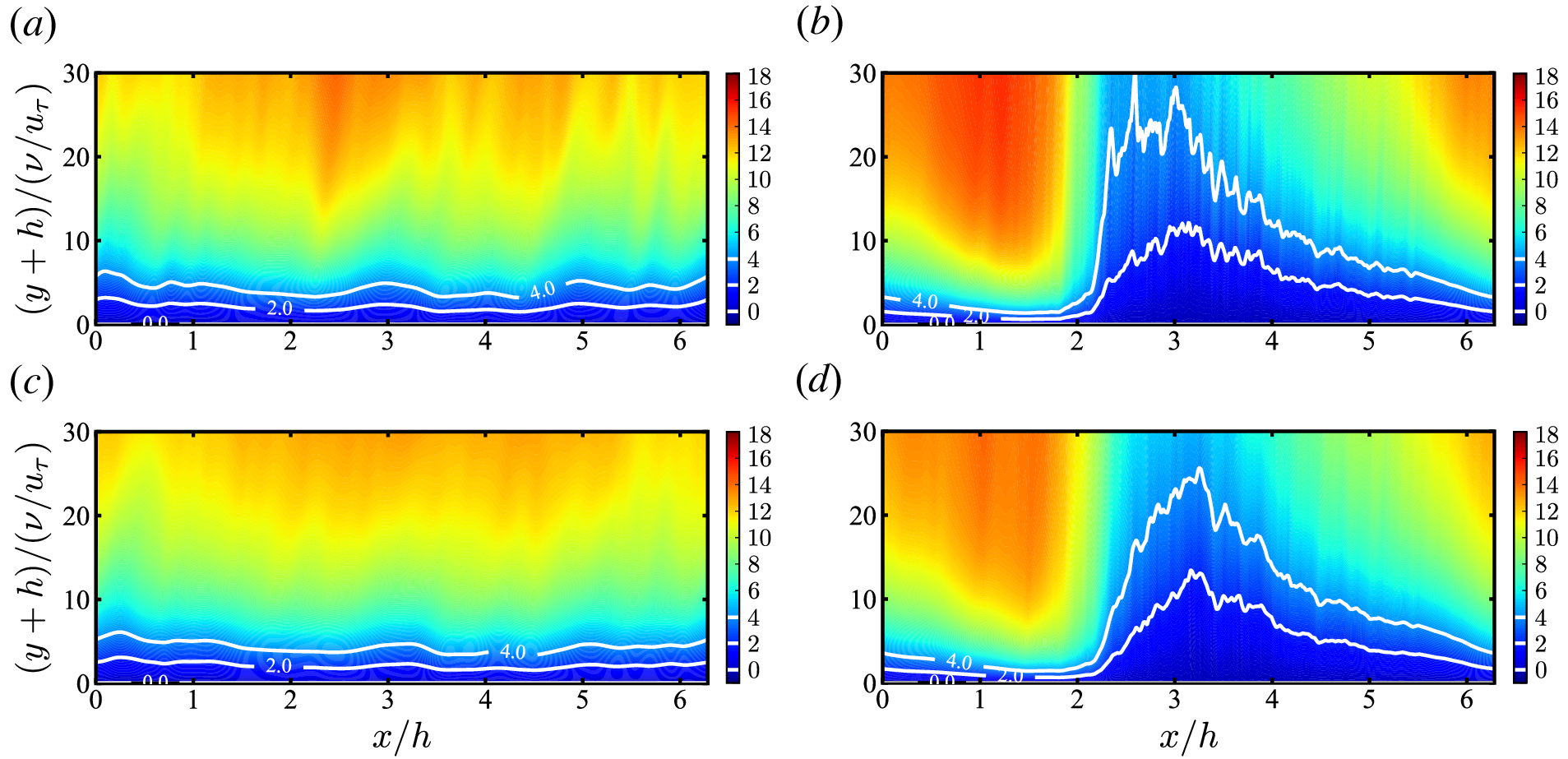}
	\end{minipage}
	\caption{\GK{Spanwise-averaged instantaneous} \REVR{\REV{(\textit{a,b})}} temperature ${\left< \theta \right>}_{z}/\theta_{\tau}$ and \REVR{\REV{(\textit{c,d})}} streamwise velocity ${\left< u \right>}_{z}/u_{\tau}$ near the lower wall at the same \GK{instant as in} figure \ref{fig:structures} \GKGK{in the permeable case \MU{($\beta u_{b}=0.5$)}}.
		The white lines indicate the isolines of ${\left< \theta \right>}_{z}/\theta_{\tau}=0$--$4$ and ${\left< u \right>}_{z}/u_{\tau}=0$--$4$.
		(\textit{a,c}) \REVGK{$Re_{b}=8\times10^{3}$, (\textit{b,d}) $Re_{b}=4\times10^{4}$.}
		\label{fig:near-wall}}
\end{figure}

%\subsection{Large-scale spanwise rolls}\label{structure}
\MU{Let us now turn to} \GK{turbulence} structures \GK{over the permeable wall}.
Instantaneous flow and thermal structures are shown \REVGK{at $Re_b=8\times 10^3$ and $Re_b=4\times 10^4$ in the permeable case \MU{($\beta u_b=0.5$)}} in figure \ref{fig:structures}.
The \GK{grey} objects represent the small-scale vortex structures identified in terms of the positive isosurfaces of the second invariant of the velocity gradient tensor \REVGK{$Q=-\left(\partial u_{i}/\partial x_{j}\right)\left(\partial u_{j}/\partial x_{i}\right)/2$},
and the \GK{dark} grey objects show the high-temperature regions, \GKGK{$\theta'>0$}.
\GK{At $Re_b=8\times 10^3$}, \MU{we detect} almost the same \GK{turbulence} structures as \GK{those} \MU{commonly} observed in wall turbulence.
The streamwise vortices %and streaks 
near the walls \GK{appear} \GK{roughly homogeneously} \MU{distributed} in the wall-parallel directions.
\REVR{\MU{On the contrary, at $Re_b=4\times 10^4$ (for which the ultimate state has been observed) very different large-scale turbulence structures in the form of spanwise-aligned rollers which are propagating downstream can be seen.}}
The \GK{colour in the figures represents \REVGK{the level of} the wall-normal velocity on the permeable walls}, exhibiting strong coherence in the spanwise direction.
The small-scale vortex structures cluster \GK{around} the blowing region, whereas \GK{high temperature concentrates} \GK{in} the suction region.
The vectors on the plane $z/h=0$ show the spanwise-averaged velocity fluctuations, $({\left< u \right>}_{z}-{\left< u \right>}_{xzt},{\left< v \right>}_{z})$, \GK{indicating} large-scale spanwise rolls with the length scale comparable with the channel half width $h$.
\GK{This} remarkable turbulence modulation originates from \GK{the} Kelvin--Helmholtz type \REVGK{of shear-layer instability} over \REVGK{a} permeable \GK{wall} \citep{Jimenez2001}.

\REVR{\REVGK{By their linear stability analyses, \cite{Jimenez2001} have shown that \MU{the mean turbulent velocity profile in plane channel flow (including background eddy viscosity)} can be unstable to infinitesimal disturbances of finite streamwise wavenumber over a permeable wall for the \MU{permeability} parameter $\beta > \beta_c$, $\beta_c$ being a critical value.
The origin of this instability has been identified as the Kelvin--Helmholtz \MU{mechanism in a free shear layer} by analytically relating an unstable %(sinuous) 
eigensolution in piecewise-linear inviscid flow over a permeable (free-slip) plane of $\beta>0$ with the eigensolution of the Kelvin--Helmholtz instability at $\beta\rightarrow\infty$.
Since the large-scale spanwise rolls in permeable-channel flow arise from the Kelvin--Helmholtz instability, they should exhibit similar properties to those of turbulence structures in free shear layers, \MU{such as a mixing layer or a jet.}
\MU{Now, in a self-similar turbulent mixing layer, large-scale spanwise vortical structures with a length scale comparable to the shear-layer thickness appear, inducing velocity fluctuations of the order of the velocity difference across the layer, such that the Taylor dissipation law holds \citep[see e.g.][]{Rogers1994}.}}
\REVGK{In the bulk region of turbulent permeable-channel flow, as in free shear layers, the large-scale rolls \MU{with a} length scale $h$ (corresponding to the free-shear-layer thickness), which undergo the velocity difference of $O(u_{b})$ (corresponding to the velocity difference across the free shear layer), can induce velocity fluctuations of $O(u_{b})$, as shown in figure \ref{fig:rms_ub_vel_te} (\REV{\textit{d,f}}).}
\GK{Accordingly}, the Taylor dissipation law \GK{$\epsilon\sim u_{b}^{3}/h$ can hold \MU{in this case},} and the total energy budget equation (\ref{eq:budget}) provides us \GK{with $c_{f}\sim Re_b^0$}.}

\REV{Figure \ref{fig:near-wall} shows the spanwise-averaged instantaneous temperature and streamwise velocity in the viscous sublayer.}
The white isolines, ${\left< u \right>}_{z}/u_{\tau}=0$--$4$ and ${\left< \theta \right>}_{z}/\theta_{\tau}=0$--$4$, indicate the thermal conduction layer and the \GK{(viscous) linear sublayer}.
\REVR{\REV{Note that the null isolines \REVGK{cannot} be observed except \REVGK{for the wall surface, implying no flow separation from the wall}.}}
\REVR{\REV{At $Re_b=4\times 10^4$}}, the temperature and velocity distributions near the wall differ greatly from those \REVR{\REV{at $Re_b=8\times 10^3$}}, and significantly large-amplitude temperature and velocity fluctuations are induced even in the close vicinity of the wall, $(y+h)/(\nu/u_{\tau})\sim 10^0$.
\GK{The near-wall low-temperature and low-velocity fluids are \REVGK{blown up} from the permeable wall, while the high-temperature and high-velocity fluids are \REVR{\REVGK{sucked towards}} the wall, \REVR{\MU{inducing events with large turbulent heat flux and Reynolds shear stress.}}
In spite of such significant enhancement of heat and momentum transfer, \REVR{\REV{there is no flow separation over the permeable wall (see figure \ref{fig:near-wall}\textit{d}) unlike \MU{in} flows over a rough wall \citep[see e.g. figure 10\textit{e} in][]{MacDonald2019a}.}
\REVGK{\MU{This is because the build-up of high-pressure is counteracted by wall-transpiration in the case of a permeable wall.}
Pressure fluctuations and resulting flow separation yield dissimilarity between heat and momentum transfer as observed in a channel with surface roughness.
In permeable-channel flow, however, \MU{the absence of flow separation implies the similarity between heat and momentum transfer.}}}
Therefore, heat transfer can \REVGK{also} be enhanced by the large-scale spanwise rolls comparably with momentum transfer, so that temperature fluctuations are of the order of $\theta_{b}$ \REVGK{(see figure~\ref{fig:rms_ub_vel_te}\textit{b}}).
As a consequence, the wall-normal heat flux scales with $u_{b}\theta_{b}$, leading to the ultimate scaling $St\sim Re_b^0$.}

\GK{
Finally, \GK{following the argument} in \cite{Kawano2021}, we would like to suggest the possibility of the ultimate state in practical applications.
\REVR{\MU{Let us consider a wall perforated with many fine holes connected to an adjacent constant-pressure plenum chamber.}
\REVGK{On such a porous wall the fluid is expected to \MU{move into or out of the wall in the wall-normal direction through the holes, implying a nearly zero  wall-parallel velocity component in the wall plane.}}}
Supposing the flow through the holes to be laminar Hagen--Poiseuille flow, the \REVR{\REV{\MU{permeability} parameter}} $\beta$ can be expressed rigorously as $\beta=d^{2}/(32\nu l)$, and the dimensionless \REVR{\MU{permeability} parameter} is \GK{given by}
\begin{eqnarray}
\label{eq:bub_hole}
\displaystyle
\beta u_{b}
\GK{=}
\frac{1}{32}{\left( \frac{d}{h} \right)}^{2}\frac{h}{l}Re_{b},
\end{eqnarray}
where $d$ and $l$ represent the diameter of the holes and the thickness of the wall, respectively.
Taking into consideration that all the pressure power on the permeable wall \GKGK{in} channel flow \GKGK{should be} consumed to drive the viscous flow in the holes, \REVR{\REVGK{the mean velocity $v_m$ in the holes would be comparable with the RMS wall-normal velocity $v_{\rm rms}$ on the permeable wall \citep{Kawano2021}.
Let us \MU{further} suppose that the thickness $l$ of the porous wall is of the order of the channel half width $h$.}}
Substitution of $l/h\sim1$ in (\ref{eq:bub_hole}) yields $d/h\sim {(\beta u_{b})}^{1/2}Re_{b}^{-1/2}$.
Thus, the porous wall with the geometry of $l/h\sim1$ and $10^{-3}\lesssim d/h\lesssim10^{-2}$ could be characterised by the \REVR{\REV{\MU{permeability} parameter}} \GK{$\beta u_{b}\sim 10^0$} at $10^{4}\lesssim Re_{b}\lesssim10^{6}$, where the ultimate state should be observed.
The mean velocity in the holes could be estimated to be $v_{m}\sim10^{-2}u_{b}$, since the RMS wall-normal velocity on the permeable wall is approximately 1\% of $u_{b}$ \REVGK{at $Re_b\sim 10^4$} for $\beta u_{b}=0.5$ (see figure~\ref{fig:rms_ub_vel_te}\textit{f}).
At $10^{4}\lesssim Re_{b}\lesssim10^{6}$ the Reynolds number of the flow in the holes, $v_{m}d/\nu\sim10^{-2}u_{b}d/\nu\sim 10^{-2}Re_{b}d/h$, is in the range $10^{0}\lesssim v_{m}d/\nu\lesssim10^{1}$, where the flow is laminar and \MU{can be expected to fulfill} \GK{the} `Darcy law'.
Therefore, we believe that the ultimate state \GK{can be achieved} in the \GK{above realistic wall-flow configuration}.
}

\vspace*{-4mm}
\section{\GKGK{Summary and outlook}}\label{sec:summary}
We have investigated turbulent heat \GK{and momentum} transfer \GKGK{numerically} in \GK{internally heated permeable-channel flow \MU{with a} constant bulk mean velocity and temperature, $u_b$ and $\theta_b$, for $Pr=1$}.
\GK{On the permeable walls at $y=\pm h$} the wall-normal velocity is \GK{assumed to be} proportional to the local pressure fluctuations, \GK{i.e.} $v(y=\pm h)=\pm\beta p/\rho$.
%The zero net mass flux \GKGK{through} the wall is instantaneously ensured, and the blowing and suction are passively driven without any additional energy input.
%The permeability parameter is set to $\beta u_{b}=0$ (the impermeable case), $\beta u_{b}=0.3$ (the \GK{less-permeable} case) and $\beta u_{b}=0.5$ (the \GK{permeable} case).
%\GK{In the less-permeable case \MU{($\beta u_{b}=0.3$)}, the Stanton number $St$ and the friction coefficient $c_{f}$ are comparable to those in the impermeable case, and we have observed the scaling $St\approx c_{f}\sim Re_{b}^{-1/4}$ consistent with the classical Blasius \GKGK{empirical law} commonly observed in wall turbulence}.

\GK{In the permeable \GKGK{channel} \MU{($\beta u_{b}=0.5$)}, we have found the \REVR{\REV{transition}} of the scaling of \GKGK{the Stanton number} $St$ and \GKGK{the friction coefficient} $c_{f}$ from the Blasius \GKGK{empirical law} $St\approx c_{f}\sim Re_{b}^{-1/4}$ to the ultimate state of $St\sim Re_b^0$ and $c_{f}\sim Re_b^0$ at the bulk Reynolds number $Re_b\sim 10^4$.}
\REVR{\REV{At $Re_b\lesssim 10^{4}$}}, there \GK{are} no significant changes in turbulence statistics or structures from the \GKGK{impermeable case \MU{($\beta u_b=0$)}}.
The ultimate state found \REVR{at \REV{$Re\gtrsim 10^{4}$}} is attributed to the appearance of large-scale \GK{spanwise rolls stemming from the Kelvin--Helmholtz type \REVGK{of} shear-layer instability over the permeable wall.}
On the permeable wall surface the blowing and suction are \GK{excited by the Kelvin--Helmholtz wave} which is \GK{roughly} uniform in the spanwise direction.
\GK{Near-wall low-temperature and low-velocity fluids are \REVGK{blown up} from the permeable wall, while the high-temperature and high-velocity fluids are \REVR{\REVGK{sucked towards}} the wall, largely producing the turbulent heat flux and the Reynolds shear stress.}
\GK{Such} remarkable turbulence modulation extends to the \GK{close} vicinity of the wall, \GKGK{$|y\pm h|/(\nu/u_{\tau})\sim10^{0}$}.
Unlike \MU{in the case of} rough walls, there is no flow separation, \GK{so that heat transfer is enhanced \MU{in a way comparable to} momentum transfer}.
\GK{The key to the achievement of the ultimate state in \REV{permeable-channel flow} is the significant heat and momentum transfer enhancement without flow separation by large-scale spanwise rolls of the length scale of $O(h)$.}
\GK{The large-scale rolls can induce the large-amplitude velocity fluctuations of $O(u_{b})$ \REVR{\REVGK{as in free shear layers} and \MU{they} \REVGK{can similarly induce the large-amplitude}} temperature fluctuations of $O(\theta_{b})$,
leading to the Taylor dissipation law $\epsilon\sim u_{b}^{3}/h$ (or equivalently $c_{f}\sim Re_b^0$) and \MU{to} the ultimate scaling $q_w/(\rho c_{p})\sim u_{b}\theta_{b}$ (or equivalently $St\sim Re_b^0$).}

\GKGK{In this \GKGKGK{study} the ultimate state has been achieved in internally heated permeable-channel flow for the \REVR{\REV{\MU{permeability} parameter}} $\beta u_b=0.5$ and the streamwise period $L_{x}=2\pi h$.
If we consider a different thermal configuration, e.g. constant temperature difference $\Delta\theta$ between the permeable walls, the same large-scale rolls appear to induce large-amplitude temperature fluctuations of $O(\Delta\theta)$, so that the ultimate scaling $q_w/(\rho c_{p})\sim u_{b}\Delta\theta$ should be achieved as well.
Concerning the dependence of the ultimate state on $\beta u_b$ and $L_{x}$, our preliminary study has shown that \MU{a} slight reduction to $\beta u_b=0.45$ delays the onset of the ultimate state until $Re_b\sim 2\times 10^4$ and that the longer \GKGKGK{$L_x=4\pi h$} can occasionally accommodate the larger streamwise wavelength of the spanwise rolls, yielding the lower onset $Re_b$ and \REVGK{the greater value of the prefactor in} the ultimate scaling.}
\GKGKGK{A detailed examination is left for a future study.}

\vspace*{-4mm}
\section*{Acknowledgements}
\REVGK{We are grateful to Professsor M. Uhlmann for his useful comments on this paper.}
This work was supported by the Japanese Society for Promotion of Science (JSPS) KAKENHI Grant Numbers 19K14889 and 18H01370.

\bibliography{ref}
\bibliographystyle{jfm}

\end{document}